\documentclass{pspum-l}
\usepackage{amsmath, amssymb, amsthm, comment, accents, slashed, url}
\usepackage{amsrefs}

\newtheorem*{conjecture}{Conjecture}
\newtheorem*{"theorem"}{``Theorem''}

\newcommand{\del}{\partial}
\newcommand{\delb}{{\bar\partial}}

\newcommand{\ket}[1]{|#1\rangle}

\newcommand{\vev}[1]{\langle #1 \rangle}

\newcommand{\ch}{\mathop{\mathrm{ch}}\nolimits}
\newcommand{\ich}{\mathit{ch}}

\newcommand{\Tr}{\mathop{\mathrm{Tr}}\nolimits}

\newcommand{\longto}{\longrightarrow}

\newcommand{\Z}{\mathbb{Z}}

\newcommand{\R}{\mathbb{R}}
\newcommand{\C}{\mathbb{C}}

\newcommand{\CP}{\mathbb{CP}}

\let\nc\newcommand
\let\renc\renewcommand
\nc{\wbar}{\overline}
\let\td\tilde
\let\wtd\widetilde
\let\wht\widehat
\let\mcl\mathcal

\nc{\ab}{{\bar{a}}} \nc{\at}{\tilde{a}} \nc{\ah}{\hat{a}}
\nc{\bb}{{\bar{b}}} \nc{\bt}{\tilde{b}} \nc{\bh}{\hat{b}}
\nc{\cb}{{\bar{c}}} \nc{\ct}{\tilde{c}} 
\nc{\db}{{\bar{d}}} \nc{\dt}{\tilde{d}} \renc{\dh}{\hat{d}}
\nc{\eb}{{\bar{e}}} \nc{\et}{\tilde{e}} \nc{\eh}{\hat{e}}
\nc{\fb}{{\bar{f}}} \nc{\ft}{\tilde{f}} \nc{\fh}{\hat{f}}
\nc{\gb}{{\bar{g}}} \nc{\gt}{\tilde{g}} \nc{\gh}{\hat{g}}
\nc{\hb}{{\bar{h}}} \nc{\hh}{\hat{h}} 
\nc{\ib}{{\bar{\imath}}} \nc{\ih}{\hat{\imath}} 
\nc{\jb}{{\bar{\jmath}}} \nc{\jt}{\tilde{\jmath}} \nc{\jh}{\hat{\jmath}}
\nc{\kb}{{\bar{k}}} \nc{\kt}{\tilde{k}} \nc{\kh}{\hat{k}}
\nc{\lb}{{\bar{l}}} \nc{\lt}{\tilde{l}} \nc{\lh}{\hat{l}}
\nc{\mb}{{\bar{m}}} \nc{\mt}{\tilde{m}} \nc{\mh}{\hat{m}}
\nc{\nb}{{\bar{n}}} \nc{\nt}{\tilde{n}} \nc{\nh}{\hat{n}}
\nc{\ob}{{\bar{o}}} \nc{\ot}{\tilde{o}} \nc{\oh}{\hat{o}}
\nc{\pb}{{\bar{p}}} \nc{\pt}{\tilde{p}} \nc{\ph}{\hat{p}}
\nc{\qb}{{\bar{q}}} \nc{\qt}{\tilde{q}} \nc{\qh}{\hat{q}}
\nc{\rb}{{\bar{r}}} \nc{\rt}{\tilde{r}} \nc{\rh}{\hat{r}}
\renc{\sb}{{\bar{s}}} \nc{\st}{\tilde{s}} \nc{\sh}{\hat{s}}
\nc{\tb}{{\bar{t}}} \renc{\th}{\hat{t}} 
\nc{\ub}{{\bar{u}}} \nc{\ut}{\tilde{u}} \nc{\uh}{\hat{u}}
\nc{\vb}{{\bar{v}}} \nc{\vt}{\tilde{v}} \nc{\vh}{\hat{v}}
\nc{\wb}{{\bar{w}}} \nc{\wt}{\tilde{w}} \nc{\wh}{\hat{w}}
\nc{\xb}{{\bar{x}}} \nc{\xt}{\tilde{x}} \nc{\xh}{\hat{x}}
\nc{\yb}{{\bar{y}}} \nc{\yt}{\tilde{y}} \nc{\yh}{\hat{y}}
\nc{\zb}{{\bar{z}}} \nc{\zt}{\tilde{z}} \nc{\zh}{\hat{z}}

\nc{\Ab}{\wbar{A}} \nc{\At}{\wtd{A}} \nc{\Ah}{\wht{A}}
\nc{\Bb}{\wbar{B}} \nc{\Bt}{\wtd{B}} \nc{\Bh}{\wht{B}}
\nc{\Cb}{\wbar{C}} \nc{\Ct}{\wtd{C}} \nc{\Ch}{\wht{C}}
\nc{\Db}{\wbar{D}} \nc{\Dt}{\wtd{D}} \nc{\Dh}{\wht{D}}
\nc{\Eb}{\wbar{E}} \nc{\Et}{\wtd{E}} \nc{\Eh}{\wht{E}}
\nc{\Fb}{\wbar{F}} \nc{\Ft}{\wtd{F}} \nc{\Fh}{\wht{F}}
\nc{\Gb}{\wbar{G}} \nc{\Gt}{\wtd{G}} \nc{\Gh}{\wht{G}}
\nc{\Hb}{\wbar{H}} \nc{\Ht}{\wtd{H}} \nc{\Hh}{\wht{H}}
\nc{\Ib}{\wbar{I}} \nc{\It}{\wtd{I}} \nc{\Ih}{\wht{I}}
\nc{\Jb}{\wbar{J}} \nc{\Jt}{\wtd{J}} \nc{\Jh}{\wht{J}}
\nc{\Kb}{\wbar{K}} \nc{\Kt}{\wtd{K}} \nc{\Kh}{\wht{K}}
\nc{\Lb}{\wbar{L}} \nc{\Lt}{\wtd{L}} \nc{\Lh}{\wht{L}}
\nc{\Mb}{\wbar{M}} \nc{\Mt}{\wtd{M}} \nc{\Mh}{\wht{M}}
\nc{\Nb}{\wbar{N}} \nc{\Nt}{\wtd{N}} \nc{\Nh}{\wht{N}}
\nc{\Ob}{\wbar{O}} \nc{\Ot}{\wtd{O}} \nc{\Oh}{\wht{O}}
\nc{\Pb}{\wbar{P}} \nc{\Pt}{\wtd{P}} \nc{\Ph}{\wht{P}}
\nc{\Qb}{\wbar{Q}} \nc{\Qt}{\wtd{Q}} \nc{\Qh}{\wht{Q}}
\nc{\Rb}{\wbar{R}} \nc{\Rt}{\wtd{R}} \nc{\Rh}{\wht{R}}
\nc{\Sb}{\wbar{S}} \nc{\St}{\wtd{S}} \nc{\Sh}{\wht{S}}
\nc{\Tb}{\wbar{T}} \nc{\Tt}{\wtd{T}} \nc{\Th}{\wht{T}}
\nc{\Ub}{\wbar{U}} \nc{\Ut}{\wtd{U}} \nc{\Uh}{\wht{U}}
\nc{\Vb}{\wbar{V}} \nc{\Vt}{\wtd{V}} \nc{\Vh}{\wht{V}}
\nc{\Wb}{\wbar{W}} \nc{\Wt}{\wtd{W}} \nc{\Wh}{\wht{W}}
\nc{\Xb}{\wbar{X}} \nc{\Xt}{\wtd{X}} \nc{\Xh}{\wht{X}}
\nc{\Yb}{\wbar{Y}} \nc{\Yt}{\wtd{Y}} \nc{\Yh}{\wht{Y}}
\nc{\Zb}{\wbar{Z}} \nc{\Zt}{\wtd{Z}} \nc{\Zh}{\wht{Z}}

\nc{\CA}{\mcl{A}} \nc{\CAb}{\wbar{\CA}} \nc{\CAt}{\wtd{\CA}} \nc{\CAh}{\wht{\CA}}
\nc{\CB}{\mcl{B}} \nc{\CBb}{\wbar{\CB}} \nc{\CBt}{\wtd{\CB}} \nc{\CBh}{\wht{\CB}}
\nc{\CC}{\mcl{C}} \nc{\CCb}{\wbar{\CC}} \nc{\CCt}{\wtd{\CC}} \nc{\CCh}{\wht{\CC}}
\nc{\CD}{\mcl{D}} \nc{\CDb}{\wbar{\CD}} \nc{\CDt}{\wtd{\CD}} \nc{\CDh}{\wht{\CD}}
\nc{\CE}{\mcl{E}} \nc{\CEb}{\wbar{\CE}} \nc{\CEt}{\wtd{\CE}} \nc{\CEh}{\wht{\CE}}
\nc{\CF}{\mcl{F}} \nc{\CFb}{\wbar{\CF}} \nc{\CFt}{\wtd{\CF}} \nc{\CFh}{\wht{\CF}}
\nc{\CG}{\mcl{G}} \nc{\CGb}{\wbar{\CG}} \nc{\CGt}{\wtd{\CG}} \nc{\CGh}{\wht{\CG}}
\nc{\CH}{\mcl{H}} \nc{\CHb}{\wbar{\CH}} \nc{\CHt}{\wtd{\CH}} \nc{\CHh}{\wht{\CH}}
\nc{\CI}{\mcl{I}} \nc{\CIb}{\wbar{\CI}} \nc{\CIt}{\wtd{\CI}} \nc{\CIh}{\wht{\CI}}
\nc{\CJ}{\mcl{J}} \nc{\CJb}{\wbar{\CJ}} \nc{\CJt}{\wtd{\CJ}} \nc{\CJh}{\wht{\CJ}}
\nc{\CK}{\mcl{K}} \nc{\CKb}{\wbar{\CK}} \nc{\CKt}{\wtd{\CK}} \nc{\CKh}{\wht{\CK}}
\nc{\CL}{\mcl{L}} \nc{\CLb}{\wbar{\CL}} \nc{\CLt}{\wtd{\CL}} \nc{\CLh}{\wht{\CL}}
\nc{\CM}{\mcl{M}} \nc{\CMb}{\wbar{\CM}} \nc{\CMt}{\wtd{\CM}} \nc{\CMh}{\wht{\CM}}
\nc{\CN}{\mcl{N}} \nc{\CNb}{\wbar{\CN}} \nc{\CNt}{\wtd{\CN}} \nc{\CNh}{\wht{\CN}}
\nc{\CO}{\mcl{O}} \nc{\COb}{\wbar{\CO}} \nc{\COt}{\wtd{\CO}} \nc{\COh}{\wht{\CO}}
\nc{\CQ}{\mcl{Q}} \nc{\CQb}{\wbar{\CQ}} \nc{\CQt}{\wtd{\CQ}} \nc{\CQh}{\wht{\CQ}}
\nc{\CR}{\mcl{R}} \nc{\CRb}{\wbar{\CR}} \nc{\CRt}{\wtd{\CR}} \nc{\CRh}{\wht{\CR}}
\nc{\CS}{\mcl{S}} \nc{\CSb}{\wbar{\CS}} \nc{\CSt}{\wtd{\CS}} \nc{\CSh}{\wht{\CS}}
\nc{\CT}{\mcl{T}} \nc{\CTb}{\wbar{\CT}} \nc{\CTt}{\wtd{\CT}} \nc{\CTh}{\wht{\CT}}
\nc{\CU}{\mcl{U}} \nc{\CUb}{\wbar{\CU}} \nc{\CUt}{\wtd{\CU}} \nc{\CUh}{\wht{\CU}}
\nc{\CV}{\mcl{V}} \nc{\CVb}{\wbar{\CV}} \nc{\CVt}{\wtd{\CV}} \nc{\CVh}{\wht{\CV}}
\nc{\CW}{\mcl{W}} \nc{\CWb}{\wbar{\CW}} \nc{\CWt}{\wtd{\CW}} \nc{\CWh}{\wht{\CW}}
\nc{\CX}{\mcl{X}} \nc{\CXb}{\wbar{\CX}} \nc{\CXt}{\wtd{\CX}} \nc{\CXh}{\wht{\CX}}
\nc{\CY}{\mcl{Y}} \nc{\CYb}{\wbar{\CY}} \nc{\CYt}{\wtd{\CY}} \nc{\CYh}{\wht{\CY}}
\nc{\CZ}{\mcl{Z}} \nc{\CZb}{\wbar{\CZ}} \nc{\CZt}{\wtd{\CZ}} \nc{\CZh}{\wht{\CZ}}

\let\eps\epsilon
\let\ups\upsilon
\let\veps\varepsilon
\let\vtht\vartheta
\let\vsgm\varsigma
\let\vphi\varphi
\let\vrho\varrho

\nc{\alphab}{\bar{\alpha}} \nc{\alphat}{\td{\alpha}} \nc{\alphah}{\hat{\alpha}}
\nc{\betab}{\bar{\beta}}   \nc{\betat}{\td{\beta}}   \nc{\betah}{\hat{\beta}} 
\nc{\gammab}{\bar{\gamma}} \nc{\gammat}{\td{\gamma}} \nc{\gammah}{\hat{\gamma}} 
\nc{\deltab}{\bar{\delta}} \nc{\deltat}{\td{\delta}} \nc{\deltah}{\hat{\delta}} 
\nc{\epsilonb}{\bar{\eps}} \nc{\epsilont}{\td{\eps}} \nc{\epsilonh}{\hat{\eps}} 
\nc{\vepsb}{\bar{\veps}}   \nc{\vepst}{\td{\veps}}   \nc{\vepsh}{\hat{\veps}} 
\nc{\zetab}{\bar{\zeta}}   \nc{\zetat}{\td{\zeta}}   \nc{\zetah}{\hat{\zeta}} 
\nc{\etab}{\bar{\eta}}     \nc{\etat}{\td{\eta}}     \nc{\etah}{\hat{\eta}} 
\nc{\thetab}{\bar{\theta}} \nc{\thetat}{\td{\theta}} \nc{\thetah}{\hat{\theta}} 
\nc{\vthetab}{\bar{\vtht}} \nc{\vthetat}{\td{\vtht}} \nc{\vthetah}{\hat{\vtht}} 
\nc{\lambdab}{\bar{\lambda}} \nc{\lambdat}{\td{\lambda}} \nc{\lambdah}{\hat{\lambda}} 
\nc{\iotab}{\bar{\iota}}   \nc{\iotat}{\td{\iota}}   \nc{\iotah}{\hat{\iota}} 
\nc{\kappab}{\bar{\kappa}} \nc{\kappat}{\td{\kappa}} \nc{\kappah}{\hat{\kappa}} 
\nc{\lmdb}{\bar{\lmd}}     \nc{\lmdt}{\td{\lmd}}     \nc{\lmdh}{\hat{\lmd}} 
\nc{\mub}{\bar{\mu}}       \nc{\mut}{\td{\mu}}       \nc{\muh}{\hat{\mu}} 
\nc{\nub}{\bar{\nu}}       \nc{\nut}{\td{\nu}}       \nc{\nuh}{\hat{\nu}} 
\nc{\xib}{\bar{\xi}}       \nc{\xit}{\td{\xi}}       \nc{\xih}{\hat{\xi}} 
\nc{\pib}{\bar{\pi}}       \nc{\pit}{\td{\pi}}       \nc{\pih}{\hat{\pi}} 
\nc{\vpib}{\bar{\vpi}}     \nc{\vpit}{\td{\vpi}}     \nc{\vpih}{\hat{\vpi}} 
\nc{\rhob}{\bar{\rho}}     \nc{\rhot}{\td{\rho}}     \nc{\rhoh}{\hat{\rho}} 
\nc{\vrhob}{\bar{\vrho}}   \nc{\vrhot}{\td{\vrho}}   \nc{\vrhoh}{\hat{\vrho}} 
\nc{\sigmab}{\bar{\sigma}} \nc{\sigmat}{\td{\sigma}} \nc{\sigmah}{\hat{\sigma}} 
\nc{\vsigmab}{\bar{\vsgm}} \nc{\vsigmat}{\td{\vsgm}} \nc{\vsigmah}{\hat{\vsgm}} 
\nc{\taub}{\bar{\tau}}     \nc{\taut}{\td{\tau}}     \nc{\tauh}{\hat{\tau}} 
\nc{\upsilonb}{\bar{\ups}} \nc{\upsilont}{\td{\ups}} \nc{\upsilonh}{\hat{\ups}} 
\nc{\phib}{\bar{\phi}}     \nc{\phit}{\td{\phi}}     \nc{\phih}{\hat{\phi}} 
\nc{\varphib}{\bar{\vphi}}   \nc{\varphit}{\td{\vphi}}   \nc{\varphih}{\hat{\vphi}} 
\nc{\chib}{\bar{\chi}}     \nc{\chit}{\td{\chi}}     \nc{\chih}{\hat{\chi}} 
\nc{\psib}{\bar{\psi}}     \nc{\psit}{\td{\psi}}     \nc{\psih}{\hat{\psi}} 
\nc{\omegab}{\bar{\omega}} \nc{\omegat}{\td{\omega}} \nc{\omegah}{\hat{\omega}} 

\nc{\Gammab}{\wbar{\Gamma}}     \nc{\Gammat}{\wtd{\Gamma}}     \nc{\Gammah}{\wht{\Gamma}}
\nc{\Deltab}{\wbar{\Delta}}     \nc{\Deltat}{\wtd{\Delta}}     \nc{\Deltah}{\wht{\Delta}}
\nc{\Thetab}{\wbar{\Theta}}     \nc{\Thetat}{\wtd{\Theta}}     \nc{\Thetah}{\wht{\Theta}}
\nc{\Lambdab}{\wbar{\Lambda}}   \nc{\Lambdat}{\wtd{\Lambda}}   \nc{\Lambdah}{\wht{\Lambda}}
\nc{\Xib}{\wbar{\Xi}}           \nc{\Xit}{\wtd{\Xi}}           \nc{\Xih}{\wht{\Xi}}
\nc{\Pib}{\wbar{\Pi}}           \nc{\Pit}{\wtd{\Pi}}           \nc{\Pih}{\wht{\Pi}}
\nc{\Sigmab}{\wbar{\Sigma}}     \nc{\Sigmat}{\wtd{\Sigma}}     \nc{\Sigmah}{\wht{\Sigma}}
\nc{\Upsilonb}{\wbar{\Upsilon}} \nc{\Upsilont}{\wtd{\Upsilon}} \nc{\Upsilonh}{\wht{\Upsilon}}
\nc{\Phib}{\wbar{\Phi}}         \nc{\Phit}{\wtd{\Phi}}         \nc{\Phih}{\wht{\Phi}}
\nc{\Psib}{\wbar{\Psi}}         \nc{\Psit}{\wtd{\Psi}}         \nc{\Psih}{\wht{\Psi}}
\nc{\Omegab}{\wbar{\Omega}}     \nc{\Omegat}{\wtd{\Omega}}     \nc{\Omegah}{\wht{\Omega}}

\makeatletter
\def\wbar{\accentset{{\cc@style\underline{\mskip12mu}}}}
\def\wbarl{\accentset{{\cc@style\mskip-2mu\underline{\mskip12mu}}}}
\let\wb@r\wbar
\let\wb@rl\wbarl
\renewcommand{\wbar}[1]{\wb@r{#1}}
\renewcommand{\wbarl}[1]{\wb@rl{#1}}
\renewcommand{\Tb}{\wbarl{T}}
\renewcommand{\Qb}{\wbarl{Q}}
\makeatother

\renewcommand{\psit}{\tilde\psi}
\renewcommand{\psib}{\bar\psi}

\title{Vanishing Chiral Algebras and H\"ohn--Stolz Conjecture}
\author{Junya Yagi}
\address{Center for Frontier Science, Chiba University, Japan}

\begin{document}
\maketitle

\begin{abstract}
  Given a two-dimensional quantum field theory with $(0,2)$
  supersymmetry, one can construct a chiral (or vertex) algebra. The
  chiral algebra of a $(0,2)$ supersymmetric sigma model is,
  perturbatively, the cohomology of a sheaf of chiral differential
  operators on a string K\"ahler manifold. However, it vanishes in
  some cases when instantons are taken into account. I will discuss
  the implication of this phenomenon for the H\"ohn--Stolz conjecture
  on the Witten genus.
\end{abstract}

\section{Introduction}

In his classic papers \cite{MR885560, MR970288}, Edward Witten
unraveled beautiful connections between elliptic genera, the geometry
of loop spaces, and supersymmetric sigma models in two dimensions.
Applying his idea to a special case, he was led to discover a new
genus, now called the Witten genus.  Let $M$ be a closed string
manifold of dimension $d$.  ($M$ is string if and only if it is spin
and $p_1(M)/2 = 0$.)  Then, the Witten genus $\varphi_W$ associates to
$M$ an integral modular form $\varphi_W(M)$ of weight $d/2$.

About a decade after Witten's work, Gerald H\"ohn and Stephan Stolz
independently arrived at the following

\begin{conjecture}[H\"ohn, Stolz \cite{MR1380455}]
  If $M$ admits a Riemannian metric of positive Ricci curvature, then
  $\varphi_W(M) = 0$.
\end{conjecture}

The aim of this paper is to explain how recent developments in the
study of supersymmetric sigma models shed new light on the
H\"ohn--Stolz conjecture in the K\"ahler case.  The key element in the
discussion is a remarkable phenomenon exhibited by certain
supersymmetric sigma models: the chiral algebras associated to these
models vanish.

The emergence of a chiral algebra is a characteristic property of
two-dimensional quantum field theories with $(0,2)$ supersymmetry.
Although this fact had been known \cite{Vafa:1989pa} for long time, it
had not been paid much attention by physicists until relatively
recently.  During the last several years, however, there were
significant advances in our understanding of the chiral algebras of
$(0,2)$ supersymmetric sigma models, thanks largely to the pioneering
works of Anton Kapustin \cite{Kapustin:2005pt} and Witten
\cite{MR2320663}.  What inspired the physicists was the theory of
chiral differential operators developed earlier in mathematics
\cite{MR1704283, MR1729362, MR1748287, MR2038198, MR1995376}.  I hope
that the results presented here give some inspirations back to
mathematicians.

We begin in the next section by reviewing how the Witten genus arises
in the context of supersymmetric sigma models, and how it is
interpreted in the language of loop space geometry.  Then, in Section
3, we restrict to the K\"ahler case and describe the construction of
the chiral algebras of $(0,2)$ supersymmetric sigma models.  We also
explain some of their properties and relation to chiral differential
operators.  In Section 4, we outline the argument for the vanishing
phenomenon and discuss its implication for the H\"ohn--Stolz
conjecture.  We end in Section 5 by raising some questions and
speculating on possible answers.

\section{Supersymmetric Sigma Model and the Witten Genus}

Consider a sigma model on a cylinder $\R \times S^1$ with target space
$M$.  This is a quantum field theory describing maps $\phi\colon \R
\times S^1 \to M$.  If we add a right-moving fermion $\psi_+$ with
values in $\phi^*TM$, then the theory has $(0,1)$ supersymmetry; that
is to say, there is a Hermitian fermionic operator $Q_+$ satisfying
\begin{equation}
  Q_+^2 = H - P, \qquad
  [Q_+, H] = [Q_+, P] = 0,
\end{equation}
where $H$ and $P$ are the generators of translations in time and
space, respectively.  The theory we consider in this paper is the
Euclidean version of this supersymmetric sigma model obtained by Wick
rotating the time coordinate $t \in \R$.

States annihilated by the supercharge $Q_+$ are called supersymmetric
states.  Since the theory is unitary, supersymmetric states can also
be characterized by the property that they have $H - P = 0$.  Due to
the topology of the space $S^1$, the momentum $P$ is quantized and
takes values in $-d/24 + \Z$.  (The shift by $-d/24$ comes from the
regularization of an infinite sum that appears in the definition
of~$P$.)

Now, cut out a finite segment of the cylinder and glue the two ends
with a twist to make a torus $\C/2\pi(\Z + \tau\Z)$.  The partition
function on the torus is computed by
\begin{equation}
  Z(M) = \Tr\bigl((-1)^{F} q^{(H + P)/2} \qb^{(H - P)/2}\bigr)
\end{equation}
with $q = e^{2\pi i\tau}$.  Here $(-1)^F$ is the fermion parity
operator, which equals $+1$ for bosonic states and $-1$ for fermionic
states.  Since $Q_+$ pairs bosonic and fermionic states outside its
kernel, $Z(M)$ receives contributions only from supersymmetric states.
So if we let $\CH$ be the space of supersymmetric states (that is, the
kernel of $Q_+$) and $\CH_n$ the subspace of $\CH$ defined by $P =
-d/24 + n$, we can write
\begin{equation}
  \label{Z}
  Z(M) = \Tr_\CH\bigl((-1)^F q^P\bigr)
       = q^{-d/24} \sum_{n = 0}^\infty q^n \Tr_{\CH_n} (-1)^{F}.
\end{equation}
This is the equivariant index of $Q_+$ with respect to the natural
circle action generated by $P$.  The Witten genus of $M$ is then given
by
\begin{equation}
  \label{WG}
  \varphi_W(M) = \eta(q)^d Z(M),
\end{equation}
where $\eta(q) = q^{1/24} \prod_{n=1}^\infty (1-q^n)$ is the Dedekind
eta function.  As the Witten genus and the partition function differ
only by a factor, we will mainly talk about the latter, which is
physically more natural to consider.

The geometric meaning of the partition function becomes clear if we
view the theory as supersymmetric quantum mechanics on the free loop
space $\CL M$ of $M$, and canonically quantize it.  After
quantization, the fermion obeys the anticommutation relation
\begin{equation}
  \{\psi^i(t,\sigma), \psi^j(t,\sigma')\} = g^{ij} \delta(\sigma - \sigma'),
\end{equation}
where $g$ is the metric on $M$.  This is a loop-space version of the
defining relation of Clifford algebra, with $\sigma$ serving as a
continuous index parametrizing the direction along the loop.  Since
the fermion acts on states, quantization identifies states with
spinors on $\CL M$ and $(-1)^{F}$ with the chirality operator.  On the
other hand, $Q_+$ is quantized as
\begin{equation}
  Q_+ = \int \! d\sigma \Bigl(\psi^i \frac{D}{D\phi^i}
      + g_{ij} \psi^i \del_\sigma \phi^j\Bigr),
\end{equation}
so it is the Dirac operator $D_{\CL M}$ on $\CL M$ plus an extra term
coupled to the Killing vector field generating rotations of loops.
Taking the limit $g \to 0$ in which the extra term drops out, we learn
that $Z(M)$ is the $S^1$-equivariant index of $D_{\CL M}$
\cite{MR970288}.  In this limit, supersymmetric states are harmonic
spinors on $\CL M$.

Taking the limit $g \to \infty$, we obtain a formula that expresses
$Z(M)$ in terms of characteristic classes of $M$.  Since $Q_+^2$
contains the potential
\begin{equation}
  \int \! d\sigma \, g_{ij} \del_\sigma \phi^i \del_\sigma \phi^j,
\end{equation}
in this limit supersymmetric states are supported mostly in the
neighborhood of the configurations such that $\del_\sigma\phi = 0$.
These are the constant loops, forming a copy of $M$ inside $\CL M$.
Such localized states can be thought of as spinors on $M$.  More
precisely, approximate supersymmetric states with $P = -d/24 + n$ can
be identified to leading order with spinors taking values in the
vector bundle $V_n$, which is the coefficient of $q^n$ in the series
\begin{equation}
  \bigotimes_{m = 1}^\infty S_{q^m}(TM)
  = 1 + qTM + q^2\bigl(TM + S^2(TM)\bigr) + \dotsb.
\end{equation}
Here $S_q(V) = 1 + qS(V) + q^2S^2(V) + \dotsb$.  On these localized
states, $Q_+$ acts as the Dirac operator $D_M$ on $M$.  Therefore we
have
\begin{equation}
  Z(M) = q^{-d/24} \sum_{n = 0}^\infty q^n \Ah(M; V_n).
\end{equation}
Mathematically, this formula defines the Witten genus through the
relation \eqref{WG}.
 
Although the last expression of $Z(M)$ is well defined for any closed
orientable manifold $M$, the underlying quantum field theory is not.
For the theory to not suffer from quantum anomalies, $M$ must be
string.  In addition, to ensure that the theory has a good ultraviolet
behavior without the problem of short-distance divergences, we should
require that the Ricci curvature of $M$ is nonnegative.%
\footnote{This is not a sufficient condition for the theory to have a
  well-defined ultraviolet limit.  If the Ricci curvature is positive,
  then the theory is expected to be asymptotically free and well
  defined.}
So these conditions are necessary for $Z(M)$ to be interpreted as the
$S^1$-equivariant index of the Dirac operator on $\CL M$.  Indeed, the
former is one of the assumptions for the H\"ohn--Stolz conjecture.
But what is the significance of the assumption that the Ricci
curvature is positive?

To understand this point, recall the Lichnerowicz theorem from
classical spin geometry.  This theorem says that there are no harmonic
spinors on a closed spin manifold with positive scalar curvature.
What would be the loop-space analog of the Lichnerowicz theorem?  A
natural definition for the scalar curvature of $\CL M$ at a loop
$\gamma \in \CL M$ would be the integral of the Ricci curvature of $M$
along $\gamma$.  Then, the positive Ricci curvature implies that the
scalar curvature of $\CL M$ is positive, which in turn would imply
that there are no harmonic spinors on $\CL M$.  If there are no
harmonic spinors, then the index of the Dirac operator vanishes and,
therefore, the Witten genus vanishes.  In short, the ``loop-space
Lichnerowicz theorem'' would imply the H\"ohn--Stolz conjecture.

\section{$(0,2)$ Supersymmetry and Chiral Algebra}

From now on we assume that $M$ is K\"ahler.%
\footnote{This is an assumption about the classical geometry of $M$.
  Quantum corrections generally destroy the closedness of the
  Hermitian form $\omega$, inducing a nonzero torsion $H = i(\delb -
  \del)\omega$.  $(0,2)$ supersymmetry requires $dH = 0$.  Such a
  geometry is called strong K\"ahler with torsion.}
The motivation for doing so is that in this case our model has $(0,2)$
supersymmetry.  This allows us to construct two important objects: the
$Q$-cohomology of states and chiral algebra.

Recall that when $M$ is K\"ahler, the Dirac operator splits into two
pieces as $D_M = \CD_M + \CD_M^*$.  The operator $\CD_M$ squares to
zero, $\CD_M^2 = 0$, and the sections of the spinor bundle admits a
$\Z$-grading such that $\CD_M$ has degree $1$.  Thus we can define the
$\CD_M$-cohomology.  Since $\{\CD_M, \CD_M^*\} = D_M^2$, a standard
argument shows that the $r$th $\CD_M$-cohomology group is isomorphic
to the space of harmonic spinors of degree $r$.

Similarly, in the K\"ahler case the supercharge splits as $Q_+ = Q +
Q^*$, and $Q$ satisfies the $(0,2)$ supersymmetry algebra:
\begin{equation}
  Q^2 = 0, \qquad
  \{Q, Q^*\} = H - P.
\end{equation}
The space of states is graded by the fermion charge, which equals
$(-1)^F$ modulo two and assigns $Q$ charge $1$.  So, as in the case of
$\CD_M$, we can define the $Q$-cohomology.  The fermion charge gives a
$\Z$-grading at the level of perturbation theory, but this is reduced
to a $\Z_{2n}$-grading nonperturbatively by instantons if $c_1(M) \neq
0$, where $2n$ is the greatest common divisor of $c_1(M)$.  (Since $M$
is spin, $c_1(M)$ is even.)  By the supersymmetry algebra, the $r$th
$Q$-cohomology group is isomorphic to the space of supersymmetric
states of charge $r$.  Hence, we can compute the Witten genus once we
know the $Q$-cohomology of states.

For the application to the H\"ohn--Stolz conjecture, however, it turns
out to be more useful to study another kind of $Q$-cohomology, namely
the $Q$-cohomology in the space of local operators, with the action of
$Q$ given by the supercommutator.  When the theory is conformally
invariant, the $Q$-cohomology of states and local operators are
isomorphic via the state-field correspondence.  This is not true in
general.

The $Q$-cohomology of local operators has two important properties.
One is that its elements vary holomorphically: if $\CO$ is $Q$-closed,
then $\del_\zb\CO$ is $Q$-exact because
\begin{equation}
  \del_\zb\CO
  = [H - P, \CO]
  = [\{Q, Q^*\}, \CO]
  = \{Q, \{Q^*,\CO]].
\end{equation}
The other is that it has a natural operator product expansion (OPE)
structure inherited from the underlying theory:
\begin{equation}
  \label{OPECA}
  [\CO_i(z)] \cdot [\CO_j(z')]
  \sim \sum_k c_{ij}{}^k(z - z') [\CO_k(z')].
\end{equation}
The coefficient functions $c_{ij}{}^k(z - z')$ are holomorphic except
at $z = z'$ where they can have poles.  The holomorphic OPE algebra
generated by the $Q$-cohomology classes is what we call the chiral
algebra of the $(0,2)$ supersymmetric theory.  The chiral algebra
forms a holomorphic sector of the theory, in which correlation
functions are holomorphic in the insertion points except where two
points coincide.  If we consider the A-model (regarding it as a
$(0,2)$ supersymmetric theory), these correlation functions generalize
Gromov--Witten invariants.

The chiral algebra of $(0,2)$ supersymmetric theory has the same
structure as the chiral (or vertex) algebra of conformal field theory,
except that the grading by conformal weight is missing.  In fact, the
former is generally not conformally invariant, certainly not for our
model if $c_1(M) \neq 0$.  Though classically our model has conformal
invariance, quantum corrections break it if the Ricci curvature of $M$
is nonzero.  Correspondingly, though classically the energy-momentum
tensor $T_{zz}$ belongs to the $Q$-cohomology, there is a $Q$-closed
local operator $\theta$ satisfying
\begin{equation}
  \label{QTt}
  [Q, T_{zz}] = \del_z\theta
\end{equation}
at the perturbative level.  If $c_1(M) = 0$, then $\theta$ is actually
$Q$-exact and one can find higher-order corrections to $T_{zz}$ that
make it $Q$-closed again.  This is not possible if $c_1(M) \neq 0$.
Therefore, the chiral algebra lacks the energy-momentum tensor in that
case.

Despite the possible absence of conformal invariance, the chiral
algebra is still graded nicely by conformal weight if we consider a
slightly different variant of the theory obtained by a ``twisting''
procedure.  After twisting, the components $T_{z\zb}$, $T_{\zb\zb}$ of
the energy-momentum tensor become $Q$-exact and vanish in the
$Q$-cohomology.  This implies that the generator $\Lb_0$ of
antiholomorphic scaling vanishes, hence the holomorphic dimension
$L_0$ is equal to the spin $L_0 - \Lb_0$ which takes integer values in
the twisted model.  Thus, $L_0$ provides an integral grading that is
protected from small quantum corrections.  Nonperturbatively, this
grading is reduced to $\Z_n$ by instantons which carry nonzero scaling
dimensions.

It was shown by Witten \cite{MR2320663} that, with this grading by
conformal weight, the chiral algebra of the twisted model is
perturbatively isomorphic to the cohomology of a sheaf $\CD^{\ich}_M$
of chiral differential operators (or $\beta\gamma$ systems) on $M$.
In particular, in the conformal case $c_1(M) = 0$, the character of
the cohomology of $\CD^{\ich}_M$ computes the partition function%
\footnote{On the torus (or any Riemann surface with trivial canonical
  bundle), the twisted and untwisted models are equivalent and have
  the same partition function.}
by the state-field correspondence:
\begin{equation}
  \ch H(M; \CD^{\ich}_M) = Z(M).
\end{equation}
By the same token, Kapustin \cite{Kapustin:2005pt} showed that the
perturbative chiral algebra of the A-model is isomorphic to the
cohomology of the chiral de Rham complex \cite{MR1704283}.  The case
that the theory has a general ``gauge bundle'' $E$ was considered by
Tan \cite{MR2302273} and found to be related to the construction of
Gorbounov et al.\ \cite{MR1748287}.  Our model corresponds to the case
$E = 0$, while the A-model is the case where $E$ is the holomorphic
tangent bundle of $M$.

Therefore, at the perturbative level, the chiral algebras of $(0,2)$
supersymmetric sigma models admit a nice description which is
relatively well understood mathematically.  Beyond perturbation theory
this is not the case any longer.  Instanton effects can be violent, so
much so that they can destroy many of the perturbative $Q$-cohomology
classes---sometimes all of them.

\section{Vanishing Theorem and the H\"ohn--Stolz Conjecture}

In \cite{MR2320663}, Witten made a remarkable prediction.  His claim
was that the chiral algebras of $(0,2)$ supersymmetric sigma models,
though perturbatively infinite-dimensional, could nevertheless vanish
nonperturbatively in the presence of instantons.  More specifically,
he predicted that this phenomenon occurred for $M = \CP^1$ because
instantons would induce the relation
\begin{equation}
  \label{Qt1}
  \{Q, \theta\} \propto 1.
\end{equation}
For $M = \CP^1$, nonperturbatively the chiral algebra is $\Z_2$-graded
by fermion charge, while the grading by conformal weight is completely
broken.  So the operator $\{Q,\theta\}$, which has charge $2$ and
weight $1$ perturbatively, has a chance to be proportional to the
identity.  Such a relation means that $1 = 0$ in the $Q$-cohomology,
hence the chiral algebra is identically zero.

Witten's prediction was confirmed by an explicit computation in
\cite{MR2415553}, where it was shown that the same conclusion also
holds for all complete flag manifolds of semisimple Lie groups $G$; $M
= \CP^1$ is the case $G = SL_2$.  In fact, we have a more general

\begin{"theorem"}[Yagi \cite{Yagi:2010tp}]
  If $M$ contains a rational curve with trivial normal bundle, then
  the chiral algebra of the $(0,2)$ supersymmetric sigma model
  vanishes.
\end{"theorem"}

Here the normal bundle $N_{C/M}$ of a rational curve $C \subset M$ is
the holomorphic vector bundle defined by the exact sequence
\begin{equation}
  0 \longto T_C \longto T_M|_C \longto N_{C/M} \longto 0.
\end{equation}
We put the word ``Theorem'' inside the quotation marks because this is
not really a rigorous theorem.  Mathematically, it may be better taken
as a conjecture.

Let me give the outline of the ``proof.''  Suppose that we have an
instanton-induced relation $\{Q, \theta\} = \CO$.  By counting how
many fermion charges and conformal weights are carried by instantons,
we can show that $\CO$ represents a $Q$-cohomology class of charge $0$
and weight $0$ perturbatively, which means that it is a function on
$M$.  (There cannot be $Q$-exact terms; $\CO$ must have
antiholomorphic weight $0$ to leading order, but there are no
perturbatively $Q$-exact local operators of charge $0$ with that
property.)  On functions $Q$ acts as the Dolbeault operator $\delb$,
so $\CO$ is a holomorphic function.  Moreover, it must be a constant
since we assume that $M$ is compact.  To show that this constant is
nonzero, we only need to find a correlation function that has
$\{Q,\theta\}$ inside and does not vanish nonperturbatively:
\begin{equation}
  \vev{\{Q,\theta\} \dotsb} \neq 0.
\end{equation}
The dots represent the insertion of additional operators.  We should
insert at least one operator that is not $Q$-closed, for otherwise the
correlation function would vanish by the $Q$-invariance of the theory.
A good choice is a function supported in the neighborhood of the
rational curve $C$.  If we insert such an operator, the correlation
function receives contributions only from the instantons wrapping $C$
and fluctuations around it.  (Instantons are holomorphic maps from the
worldsheet to $M$.)  Under the assumption on the normal bundle, the
fluctuations in the directions normal to $C$ make trivial
contributions and can be ignored.  Then the computation reduces to the
case of $M = \CP^1$, but we know that $\{Q,\theta\} \propto 1$ in that
case, so we can find additional operators that make the correlation
function nonzero.  Thus, we again have the relation $\{Q,\theta\}
\propto 1$, the equation $1 = 0$ holds in the $Q$-cohomology, and the
chiral algebra vanishes nonperturbatively.

Now, with this vanishing theorem in hand, what can we say about the
H\"ohn--Stolz conjecture?  The crucial observation is the following:
the $Q$-cohomology of states vanishes if the chiral algebra does.  The
argument is simple.  Notice that the $Q$-cohomology of states is
naturally a module over the chiral algebra.  If the chiral algebra
vanishes, then $[1] = 0$ and for any $Q$-closed state $\ket{\Psi}$ we
have
\begin{equation}
  [\ket{\Psi}]
  = [1] \cdot [\ket{\Psi}]
  = 0.
\end{equation}
So any $Q$-closed state is $Q$-exact, in other words, the
$Q$-cohomology of states is zero.  Recalling that the $Q$-cohomology
classes are in one-to-one correspondence with the supersymmetric
states, we conclude that the vanishing chiral algebra implies that
there are no supersymmetric states.  It follows that the Witten genus
vanishes.

We can actually say more.  There is a K\"ahler structure on $\CL M$
induced from that on $M$, so we can write $D_{\CL M} = \CD_{\CL M} +
\CD_{\CL M}^*$ as in the finite-dimensional case.  If we canonically
quantize the theory and compare the expressions of $Q$ and $\CD_{\CL
  M}$, we realize that the two operators are related by
\begin{equation}
  Q = e^{\CA_0} \CD_{\CL M} e^{-\CA_0},
\end{equation}
where the functional $\CA_0\colon \CL M \to \R$ is the symplectic
action $\CA_H$ with $H = 0$.  This shows that the $Q$-cohomology of
states is the same as the $\CD_{\CL M}$-cohomology, and the
supersymmetric states are in one-to-one correspondence with the
harmonic spinors on $\CL M$.  Therefore, if the chiral algebra
vanishes, there are no harmonic spinors on $\CL M$ either.  We should
not take this statement too seriously, however.  We do not even know
how to construct the Dirac operator on loop space yet.

\section{Questions and Speculations}

A couple of questions naturally arise.  In view of the H\"ohn--Stolz
conjecture, an obvious one would be: does the chiral algebra vanish
whenever $c_1(M) > 0$?  If true, this implies that there are no
harmonic spinors on $\CL M$ when $M$ admits a positive Ricci
curvature, proving in the K\"ahler case the ``loop-space Lichnerowicz
theorem'' and hence the conjecture.  Although it is hard to tell at
the moment whether this should be the case, we can hope that a deeper
understanding of the physics of $(0,2)$ supersymmetric sigma models
will eventually lead to a decisive answer.  As the chiral algebra is
invariant under the renormalization group, it may be helpful to study
the low-energy descriptions of these models.

Another important question is: how can we put all the arguments
presented here on a mathematically sound footing?  To this end we will
need, at the very least, a rigorous construction of the chiral
algebra.  At the perturbative level, we know that the chiral algebra
is given by the cohomology of a sheaf of chiral differential operators
on $M$.  In this approximation we only consider the neighborhood of
the constant maps which make up a copy of $M$ in the field
configuration space.  Beyond perturbation theory, we need to take into
account the contributions from all instantons.  Then, we expect that
the chiral algebra is nonperturbatively given by the cohomology of
some sheaf defined on the whole instanton moduli space, such that it
reduces on the constant maps to the sheaf of chiral differential
operators.  It will be very interesting if such a sheaf is constructed
and found to reproduce the vanishing theorem achieved through the
physical argument.


\end{document}